\documentclass{nime-alternate}
\usepackage{amsmath}
\usepackage[hidelinks]{hyperref}

\begin{document}
\conferenceinfo{NIME'18,}{June 3-6, 2018, Blacksburg, Virginia, USA.}

\title{Composing an Ensemble Standstill Work for Myo and Bela}

\numberofauthors{3}
\author{
\alignauthor
Charles P. Martin \\
       \affaddr{RITMO, Department of Informatics, University of Oslo}\\
       \email{charlepm@ifi.uio.no}
\alignauthor
Alexander Refsum Jensenius\\
       \affaddr{RITMO, Department of Musicology, University of Oslo}\\
       \email{a.r.jensenius@imv.uio.no}
\alignauthor Jim Torresen\\
       \affaddr{RITMO, Department of Informatics, University of Oslo}\\
       \email{jimtoer@ifi.uio.no}
}

\maketitle
\begin{abstract}
  This paper describes the process of developing a standstill performance work using the Myo gesture control armband and the Bela embedded computing platform.
  The combination of Myo and Bela allows a portable and extensible version of the standstill performance concept while introducing muscle tension as an additional control parameter.
  We describe the technical details of our setup and introduce Myo-to-Bela and Myo-to-OSC software bridges that assist with prototyping compositions using the Myo controller.
\end{abstract}



\section{Introduction}\label{sec:introduction}

The Myo armband from Thalmic Labs Inc.\ is a commercial gesture control interface that includes a 9-axis inertial measurement unit (IMU), as well as 8 electromyograph (EMG) sensors, in a convenient battery-powered package. 
The device communicates wirelessly via Bluetooth Low Energy (BLE). 
The Myo is typically worn on a user's lower arm, and the EMG sensors pick up signals from muscles connected to the hand and fingers. 
Continuous data streams from the  EMG and IMU sensors can be used for interactive applications, including music performance \cite{Nymoen:2015aa}.

\begin{figure}
\centering
\includegraphics[width=\columnwidth]{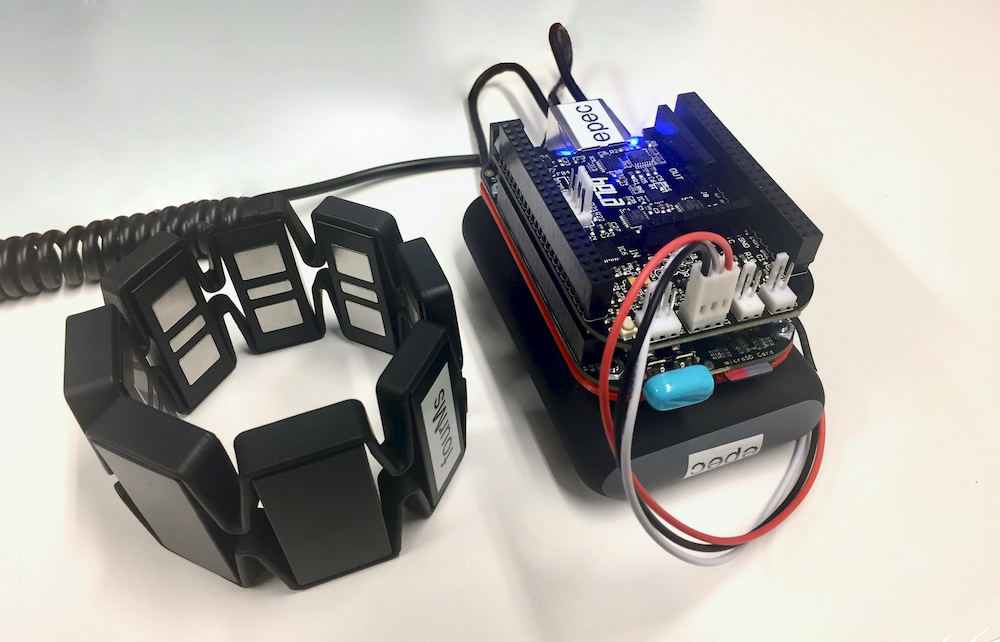}
\caption{Our mobile performance setup using a Bela and Myo, powered by a USB battery.}
\label{fig:bela-myo-setup}
\end{figure}

Recent work explored the potential of the Myo to support `microinteraction'---the use of the smallest possible control actions---in digital musical performance \cite{Jensenius:2017aa}. We have extended that preliminary work to produce a fully mobile setup, using the the Bela embedded audio computing platform~\cite{Moro:2016aa} (Fig.~\ref{fig:bela-myo-setup}), for performing \emph{Stillness Under Tension}, featuring four Myo performers.

Since its commercial release in 2015, the Myo has proven to be popular for performative interactions among electronic artists, musicians, and dancers (e.g., \cite{Nymoen:2015aa, Benson:2016aa, Di-Donato:2017ab}). However, in our work with the Myo, enthusiasm has been tempered by two technical limitations: 
First, Thalmic Labs' Myo Connect, which is often used by third-party software, does not support Linux, so it cannot be used on embedded platforms such as Raspberry Pi and Beaglebone. 
Secondly, the BLE dongle that ships with the Myo can generally only handle three simultaneous connections, requiring multiple computers for ensemble performance.

In this paper we describe our solutions to these problems. 
We have implemented a custom Bela render file that connects Bela to the Myo, as well as a pure-Python Myo-to-OSC bridge that allows composition with Myos on desktop Linux as well as other platforms.  
By connecting Myos to Belas, we work around the limitations of holding multiple BLE connections by simply using one Bela per performer.
In Section \ref{sec:performance} we discuss how these setups are applied in our standstill ensemble work for Myo.

\section{System Design}\label{system-design}

In this section we describe our hardware and software workflow for composing works using Myo armbands with the Bela embedded audio system.

\subsection{Myo-to-Bela in C++}

Working with Myo interfaces on the Bela platform is complicated by the fact that Linux operating systems are not officially supported. 
However, the BLE specification for the Myo is openly available\footnote{Myo Bluetooth Specification: \url{https://github.com/thalmiclabs/myo-bluetooth}} and several community-developed Myo libraries have emerged. 
For integration with Bela, we used \texttt{MyoLinux}\footnote{\url{https://github.com/brokenpylons/MyoLinux}}, a C++ library for Myo that uses the virtual serial port provided by the Myo's included BLE dongle. 

We implemented a custom render callback (\texttt{render.cpp} file) based on the Bela's \texttt{libpd}-enabled render file that establishes a Myo connection and sets the Myo to stream EMG and IMU data. 
Functions are included to process the incoming accelerometer and gyroscope vectors and quaternions, and produce Euler rotation angles and magnitudes of acceleration and rotation that are sent to outlets in the Pure Data patch. 
Our render file and installation instructions for the \texttt{MyoLinux} library are freely available.\footnote{Bela Myo Example repository: \url{https://doi.org/10.5281/zenodo.1216171}}

\subsection{Myo-to-OSC in Python}

In parallel with our efforts to connect Myo to Bela, we have also developed a pure-Python solution for connecting Myo armbands to OSC applications, particularly on Linux, although other platforms are also supported. 
Our \texttt{Myo-to-OSC}\footnote{Myo-to-OSC Git Repository: \url{https://doi.org/10.5281/zenodo.1216169}
} application builds on previous Python Myo libraries but adds a command line interface for listing the MAC addresses and names of available Myos, and processes the Myo's data into useful parameters, such as the Euler angles and movement magnitudes listed above. 
We use \texttt{Myo-to-OSC} to test performance interactions with a Myo before transferring to the Bela platform.

\subsection{Performance Setup with Myo and Bela}

Our performance setup includes a self-contained unit for each performer, each consisting of a Bela with the attached BLE dongle and the wireless Myo armband (shown in Fig.~\ref{fig:bela-myo-setup}). 
The Bela receives power from a regular 3000 mAh USB battery pack, which usually lasts for around 2 hours of performance. 
An important element here is that that the Bela can boot directly into a pre-configured patch. Thus a performer can turn on the power, wait 40 seconds for the boot sequence to load the patch, and be ready to start performing. 
This process appears to be reliable and we have not experienced any failures to load the patch or connect to the Myos.
For sound rendering, we have used the Bela's built-in line output to connect to headphones for personal practice, using battery-powered speakers for rehearsal, and connecting to a PA mixer for performance. 

\section{Stillness Under Tension}\label{sec:performance}

\begin{figure}
\centering
\includegraphics[width=\columnwidth]{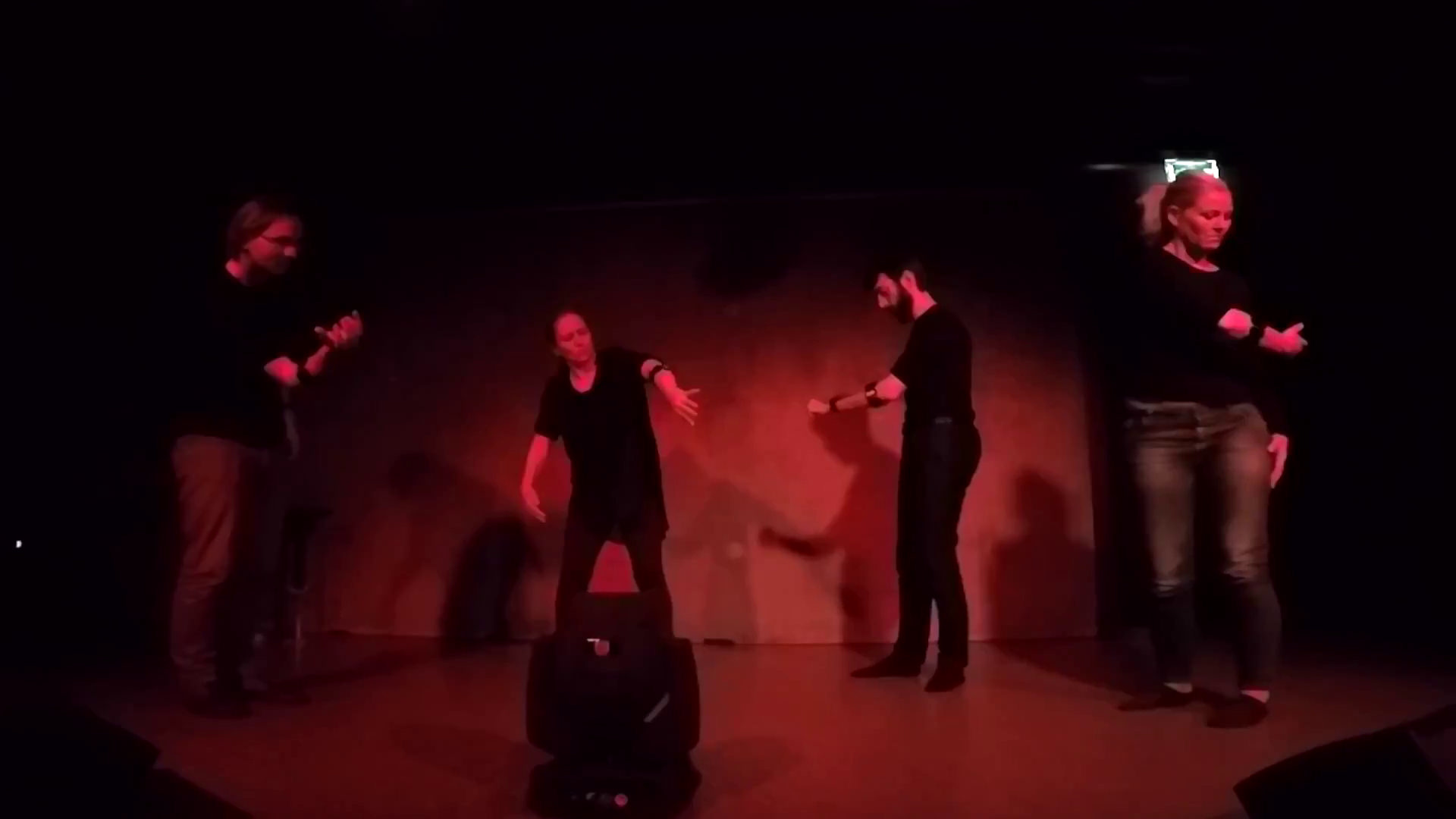}
\label{fig:sverm-myo-performance}
\caption{
The premiere of \emph{Stillness Under Tension}. 
Each performer could use their arm muscles to activate different oscillators, while their arm direction selected a different timbre and base frequency.
}\label{fig:performance}
\end{figure}

Our composition was inspired by the MicroMyo demonstration~\cite{Jensenius:2017aa}, with an additional focus on the potential for sonic variety and interaction within an ensemble. 
Sound generation was from eight sine oscillators, with the amplitude of each being modulated from the signal of one EMG sensor. 
Performers can then `select' a mix of the eight tones by tensing different muscles in their arm, for example, by holding their wrist in different positions. 
The base frequency was selected using the pitch (angle) of the Myo, the spread of the other frequencies were controlled by the yaw angle, and the roll angle controlled the gain of a distortion effect. The performance is shown in Fig. \ref{fig:performance} and a video is available.\footnote{\url{http://doi.org/10.5281/zenodo.1215956}}

The composition used an `inverse' mapping from action to sound \cite{lesaffre_sonic_2017}, in which less motion from the performer resulted in more sound. A \textit{quantity of motion} (QoM) measure was defined as the sum of the magnitudes of acceleration and rotation.
If this QoM rose above a threshold, the master amplitude was set to zero with sounds returning over 30 seconds of stillness.
This inverse mapping defines the `standstill' nature of the composition; performers must hold their arm as still as possible in order to produce sound. 

The performers were able to activate different oscillators by tensing the muscles in their arm while standing still. This required significant focus to accomplish without moving. 
As each performer had a different arm position, the oscillators had different settings, resulting in a kind of harmonic variation between the performers' sounds.

To structure performances of \emph{Stillness Under Tension}, the performers alternated between different standstill poses. They would eventually be able to explore sounds with their micromotions, while larger motion would immediately silence their Bela without disrupting the stillness of the other performers. 
The group agreed to assume around four poses over 9 minutes to enable each performer to explore a variety of sounds and to facilitate contrasts between the sonic possibilities of the composition.

\subsection*{Acknowledgements}

Partially supported by the Research Council of Norway through the EPEC (240862) and MICRO (250698) projects, and its Centres of Excellence scheme (262762).

\bibliographystyle{abbrvurl}
\bibliography{references-short}

\begin{thebibliography}{1}

\bibitem{Benson:2016aa}
C.~Benson, B.~Manaris, S.~Stoudenmier, and T.~Ward.
\newblock Sound{M}orpheus: A myoelectric-sensor based interface for sound
  spatialization and shaping.
\newblock In {\em Proc. Int. Conf. NIME}, pages 332--337, Brisbane, Australia,
  2016.

\bibitem{Di-Donato:2017ab}
B.~Di~Donato, J.~Dooley, J.~Hockman, J.~Bullock, and S.~Hall.
\newblock Myospat: A hand-gesture controlled system for sound and light
  projections manipulation.
\newblock In {\em Proc. ICMC}, Shanghai, China, 2017.

\bibitem{lesaffre_sonic_2017}
A.~R. Jensenius.
\newblock Sonic {Microinteraction} in ``the {Air}".
\newblock In M.~Lesaffre, P.-J. Maes, and M.~Leman, editors, {\em The
  {Routledge} {Companion} to {Embodied} {Music} {Interaction}}, pages 431--439.
  Routledge, New York, 2017.

\bibitem{Jensenius:2017aa}
A.~R. Jensenius, V.~G. Sanchez, A.~Zelechowska, and K.~A.~V. Bjerkestrand.
\newblock Exploring the {M}yo controller for sonic microinteraction.
\newblock In {\em Proc. Int. Conf. NIME}, pages 442--445, Copenhagen, Denmark,
  2017.

\bibitem{Moro:2016aa}
G.~Moro, A.~Bin, R.~H. Jack, C.~Heinrichs, and A.~P. e.~a. McPherson.
\newblock Making high-performance embedded instruments with {B}ela and {Pure
  Data}.
\newblock In {\em Int. Conf. on Live Interfaces}, Sussex, UK, 2016.

\bibitem{Nymoen:2015aa}
K.~Nymoen, M.~R. Haugen, and A.~R. Jensenius.
\newblock Mu{M}yo - evaluating and exploring the {M}yo armband for musical
  interaction.
\newblock In {\em Proc. Int. Conf. NIME}, pages 215--218, Baton Rouge,
  Louisiana, USA, 2015.

\end{thebibliography}
\end{document}